\documentclass{appolb}

\usepackage{graphicx}
\usepackage{amsmath}
\usepackage{bm}
\usepackage{cite}
\usepackage{soul}

\usepackage[colorlinks=true,linktocpage=true,linkcolor=blue,citecolor=blue]{hyperref}

\def \a {\alpha}
\def \b {\beta}

\def \e {\varepsilon}

\def \r {\rho}
\def \l {\lambda}
\def \m {\mu}
\def \n {\nu}
\def \s {\sigma}

\def \D {\Delta}

\def \L {\Lambda}
\def \CA {{\cal A}}

\begin{document}
\title{Baryonic spin Hall effects in Au+Au collisions at $\sqrt{s_{NN}} = 7.7 - 200$~GeV %
\thanks{Presented at Quark Matter 2022}%
}

%\author{${\rm Baochi \; Fu}^{1,2,3}$, ${\rm Shuai \; Y. \; F. \; Liu}^{4}$, ${\rm Longgang \; Pang}^{5}$, ${\rm Huichao \; Song}^{1,2,3}$, ${\rm and \; Yi \; Yin}^{4,6}$
\author{${\rm Baochi \; Fu}^{1,2,3}$, ${\rm Longgang \; Pang}^{4}$, ${\rm Huichao \; Song}^{1,2,3}$, ${\rm and \; Yi \; Yin}^{5,6}$
\address{${}^1$ Center for High Energy Physics, Peking University, Beijing 100871, China}
\address{${}^2$ Department of Physics \& State Key Laboratory of Nuclear Physics and Technology, Peking University, Beijing 100871, China}
\address{${}^3$ Collaborative Innovation Center of Quantum Matter, Beijing 100871, China}
\address{${}^4$ Key Laboratory of Quark \& Lepton Physics (MOE) and Institute of Particle Physics, Central China Normal University, Wuhan 430079, China}
\address{${}^5$ Quark Matter Research Center, Institute of Modern Physics, Chinese Academy of Sciences, Lanzhou 730000, China}
\address{${}^6$ University of Chinese Academy of Sciences, Beijing 100049, China}}

\maketitle
\begin{abstract}

In this proceeding, we present our recent prediction on the local net Lambda polarization to search for the baryonic spin Hall effect (SHE) at RHIC BES energies.  
The baryonic SHE is induced by the gradients of baryon chemical potential, which leads to local polarization separation between baryons and anti-baryons. 
Based on hydrodynamic simulations with spin Cooper-Fryer formula, we propose to use $P^{{\rm net}}_{2,y}$ and $P^{{\rm net}}_{2,z}$, the second Fourier coefficients of net spin polarization to quantify this baryonic SHE. Future experimental observation of their non-trivial signatures could strongly support the existence of the baryon SHE in hot and dense QCD matter.

\end{abstract}

%we demonstrate the baryonic SHE shows sizeable contribution to $\L$ local polarization with decreasing collision energy.

\section{Introduction}

In relativistic heavy-ion collisions, the produced quark-gluon plasma (QGP) carries a large amount of orbital angular momentum, which induces the spin polarization of quarks and final hadrons through the spin-orbital coupling~\cite{Liang:2004ph, Liang:2004xn}. 
Within the hydrodynamic approach with equilibrium assumption, the thermal vorticity developed during the system evolution leads to the hyperon spin polarization at the freeze-out surface described by the spin Cooper-Frye formula~\cite{Becattini:2013fla}. 
The related model calculations successfully describe the global $\Lambda$ polarization but fail to reproduce its azimuthal angle dependence, which is  known as the ``local spin polarization puzzle"~\cite{Becattini:2017gcx, Fu:2020oxj, Xia:2018tes, Wei:2018zfb}.
Recently, it was found that, beside the widely studied thermal vorticity effect, the spin polarization can also be induced by the shear stress tensor~\cite{Liu:2021uhn, Fu:2021pok, Becattini:2021suc, Becattini:2021iol}. 
Hydrodynamic simulations demonstrate that such shear induced polarization (SIP) is always opposite to the thermal vorticity effect and could lead to the right sign of local polarization as experimental data in some certain freeze-out conditions \cite{Fu:2021pok, Becattini:2021iol}.

If we focus on the collisions at RHIC BES energies, another spin polarization mechanism induced by the gradient of baryon chemical potential ($\nabla \mu_B$-IP) becomes relevant. 
Such effect is also called as the baryonic spin Hall effect, which can be derived from the quantum kinetic theory~\cite{Son:2012zy,Hidaka:2016yjf,Hidaka:2017auj,Hattori:2019ahi,Liu:2021uhn,Yi:2021ryh}, but have not been fully explored phenomenally. 
In this proceeding, we will briefly introduce our recent prediction on the second Fourier coefficients of net spin polarization to search the baryonic spin Hall effect (SHE)
at RHIC BES energies. For more details, see our recent work~\cite{Fu:2021pok, Fu:2022myl}.

\section{Method}
In a many-body system of fermions, the spin polarization vector is described by the axial Wigner function $\CA^{\mu}(x,p)$.
To the first order, $\CA^{\mu}$ can be expressed by the gradients of temperature $T$, flow velocity $u^\m$ and baryon chemical potential $\m_B$~\cite{Liu:2021uhn,Liu:2020dxg} (see also Refs.~\cite{Hidaka:2017auj,Yi:2021ryh}):
\begin{equation}
\begin{aligned}
{\CA}^\mu (x, p) = &\b f_0(x,p) (1 - f_0(x,p))  \e^{\m\n\a\r} \\
                   &\times \Big(
                   \underbrace{\frac{1}{2_{}} p_\n \partial_{\a}^\perp u_\r}_{\text{vorticity}}
                 - \underbrace{\frac{1}{T} u_\n p_{\a} \partial_\r T }_{\text{T-gradient}}
                 - \underbrace{\frac{p^2_\perp}{\e_0}  u_\n Q^\l_\a \s_{\r\l}}_{\text{SIP}}
                 - \underbrace{\frac{q_{B}}{\e_0 \b} u_\n p_{\a} \partial_{\r}(\b \m_B)}_{\text{baryonic SHE}}
                   \Big),
\end{aligned}
\label{eq:spin}
\end{equation}
where $ f_0(x,p) = (e^{(\e_0 - q_{B} \m_B)\beta} + 1)^{-1}$ is the Fermi-Dirac distribution function with $\e_0 = p \cdot u$, $q_B$ is the baryon number and $\beta=1/T$ is the inverse temperature.
Here, the transverse projection is defined as $\partial^\m_\perp \equiv \D^{\m\n} \partial_\n$ and $ p_{\perp}^{\mu}\equiv \D^{\mu\nu}p_{\nu}$ with $\D^{\m\n} = g^{\m\n} - u^\m u^\n$. After combining the first two terms with the ideal hydrodynamic equation $(u\cdot \partial) u_\a=-\beta^{-1}\partial^{\perp}_{\a}\beta+{\cal O}(\partial^{2})$, we will get the widely known ``thermal vorticity" term $\frac{1}{2\beta} p_\nu \partial_\a (\beta u_\rho)$.

The third term of Eq.~\ref{eq:spin} denotes the shear induced polarization, where the generalized momentum quadrupole tensor and shear stress tensor are defined as $Q^{\m\n} = -p^\m_\perp p^\n_\perp / p^2_\perp + \D^{\m\n} /3$ and $\s^{\m\n} = \partial^{(\m}_\perp u^{\n)}_{{}} - (1/3)\D^{\m\n} (\partial\cdot u)$. This expression can be obtained from chiral kinetic theory or linear response theory~\cite{Liu:2021uhn, Fu:2021pok}. In~\cite{Becattini:2021suc,Becattini:2021iol}, the statistical method also gives a similar form. Such shear term induces similar differential polarization $P_z(\phi)$ and $P_y(\phi)$ \cite{Fu:2021pok, Becattini:2021iol}  as observed in experiment, which is essential to solve the ``local spin polarization puzzle", see \cite{Yi:2021ryh, Ryu:2021lnx, Florkowski:2021xvy, Sun:2021nsg, Alzhrani:2022dpi} for recent developments of the phenomenological study.  The last term of Eq.~\ref{eq:spin} is the baryon spin Hall effect(SHE), which is induced by the gradients of baryon chemical potential and becomes significant at RHIC BES energies. 

%Noting that Eq.~\ref{eq:spin} includes all the possible contributions from the leading order hydrodynamic gradients, namely ``vorticity", ``temperature gradient", ``shear" and ``baryonic SHE".  Here, the new effects ``shear" and ``baryonic SHE" will be discussed in the following proceeding.

To calculate the differential $\L$ spin polarization $P^\mu(p)$, we employ the ``spin Cooper-Frye" formula that average $\CA^\mu(x,p)$ over the freeze-out hyper-surface:
\begin{equation}
    P^\m(\bm{p}) = \frac{\int{d \Sigma^\a p_\a \CA^\m(x,\bm{p};m) }} {2m \int{d \Sigma^\a p_\a f_0(x,p)}}\, .
\label{eq:frz}
\end{equation}
Here we consider two scenarios ``Lambda equilibrium" and ``strange memory" for the spin polarization at freeze-out, which corresponds two extreme conditions of the spin relaxation time during hadronic evolution \cite{Becattini:2013fla, Liang:2004ph, Fu:2021pok}. In the model calculation, we set $m= 1.116 \;\textrm{GeV}, q_{B}=\pm 1$ for the ``Lambda equilibrium" scenario and $m = 0.3 \;\textrm{GeV},q_{B}= \pm 1/3$ in Eq.~\eqref{eq:frz} for the ``strange memory" scenario respectively.

\begin{figure}[htb]
\centerline{%
\includegraphics[width=1.0\textwidth]{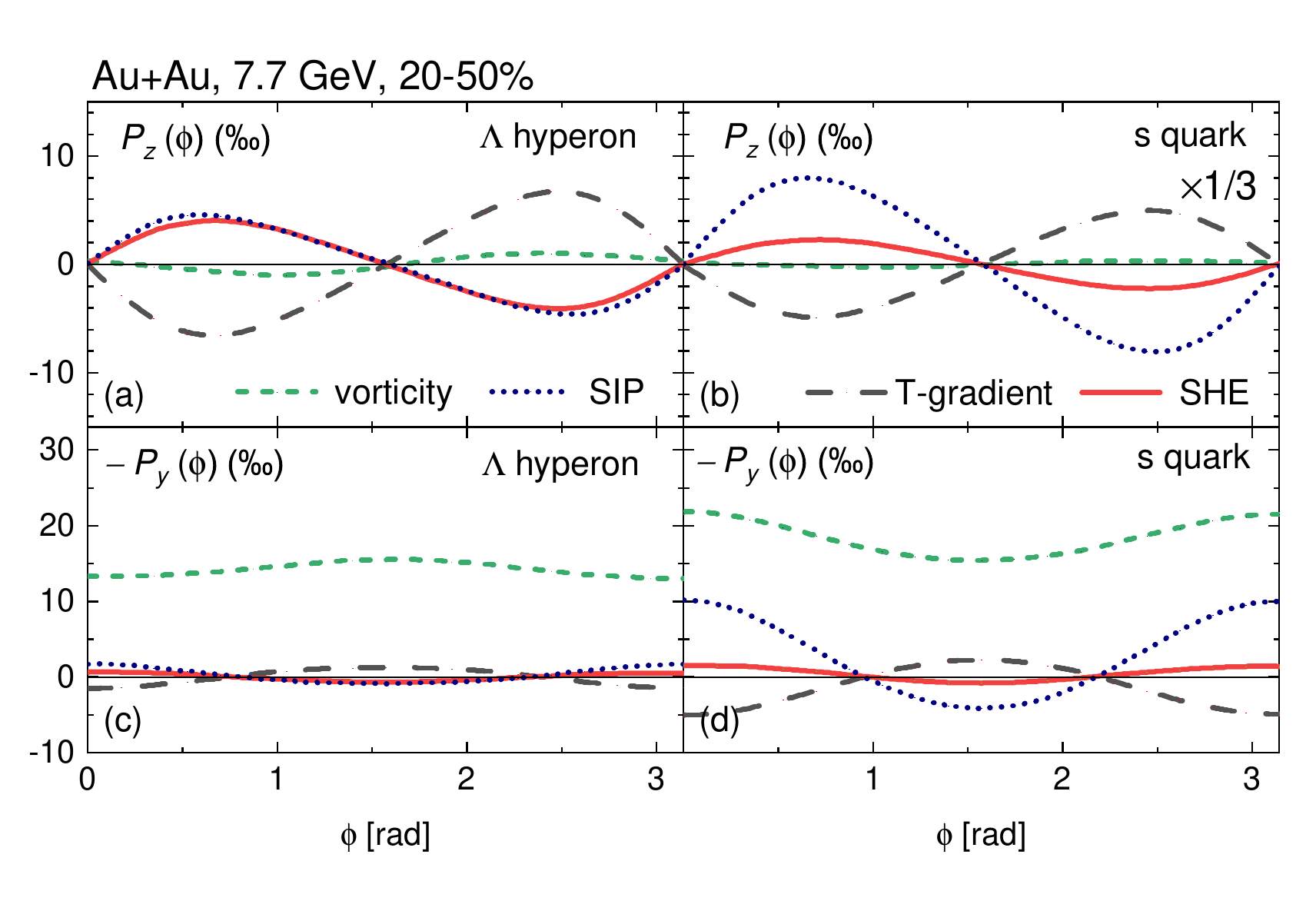}}
\caption{Differential spin polarization $P_z(\phi)$ and $-P_y(\phi)$ in 7.7 GeV Au+Au collisions from ``Lambda equilibrium" scenario and ``strange memory" scenario, respectively. The colored curves show the contributions from vorticity, T gradient, shear, and baryonic SHE.}
\label{Fig:diff}
\end{figure}

\section{Results and discussions}
In this work, we implement 3+1-d hydrodynamics MUSIC with AMPT initial conditions~\cite{Schenke:2010nt, Lin:2004en,Xu:2016hmp} to simulate Au+Au collisions at $\sqrt{s_{NN}} = 7.7 - 200$~GeV. The spin Cooper-Frye formula Eq.~\ref{eq:frz} is implemented on the hyper-surface with constant energy density, which roughly matches the chemical freeze-out temperature from the statistical model. For more details of the model and parameter set-ups please refer to \cite{Fu:2020oxj}.

Fig.~\ref{Fig:diff} shows the contributions from the vorticity, T-gradient, shear and baryonic SHE 
to the differential spin polarization in Au+Au collisions at $\sqrt{s_{NN}} = 7.7$~GeV.
Analogous to the results of high energy collisions \cite{Fu:2021pok, Becattini:2021iol, Yi:2021ryh}, the shape of $P_z(\phi)$ and $P_y(\phi)$ is mostly determined by the competition between the T-gradient and SIP effects. In the strange memory scenario, the SIP becomes more important due to the smaller mass of the spin carrier.
On the other hand, with the large chemical potential, the baryonic SHE shows significant contribution to differential polarization, especially for $P_z(\phi)$ in both scenarios at 7.7 GeV. For baryons and anti-baryons, since the sign of the baryonic SHE is opposite, a sizeable separation between their differential polarization are anticipated \cite{Fu:2022myl}.

\begin{figure}[htb]
\centerline{%
\includegraphics[width=1.0\textwidth]{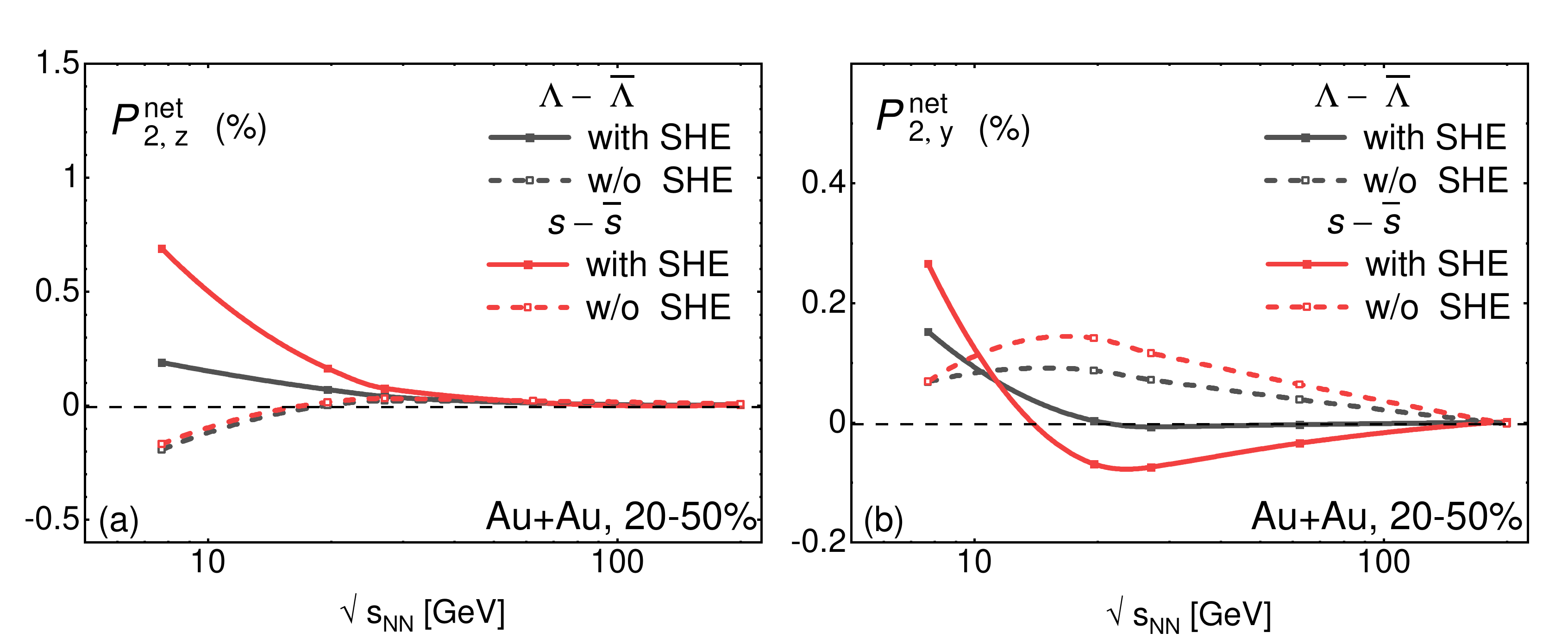}}
\caption{The second Fourier coefficients of net spin polarization, $P^{{\rm net}}_{2,z}$ and $P^{{\rm net}}_{2,y}$ as a function of collision energy in ``Lambda equilibrium" and ``strange memory" scenarios.}
\label{Fig:SHE}
\end{figure}

To characterize this separation and probe the baryon SHE signal, we construct the second harmonic component of polarization for net lambda/strangeness:
\begin{align}
\label{Obervables}
    P^{{\rm net}}_{2,z}\equiv \langle P^{{\rm net}}_{z}(\phi)\sin2\phi\rangle\, ,
    P^{{\rm net}}_{2,y}\equiv -\langle P^{{\rm net}}_{y}(\phi)\cos2\phi\rangle\, ,
\end{align}
where $P^{{\rm net}}_{z,y}(\phi) \equiv P_{z,y}(\phi)-\overline{P}_{z,y}(\phi)$ denotes the net spin polarization with 
$P_{z,y}(\phi)$ and $\overline{P}_{z,y}(\phi)$ denotes the differential polarization along longitudinal direction (z) and out-of-plane direction (y) for baryons and anti-baryons, respectively. Here  $\langle...\rangle$ is the average over the azimuthal angle.
Fig.~\ref{Fig:SHE} shows $P^{{\rm net}}_{2,z}$ and $P^{{\rm net}}_{2,y}$ in Lambda equilibrium and strange memory scenario.

When including SHE, the longitudinal component $P^{{\rm net}}_{2,z}$ increases with the decrease of collision energy, corresponding to the increasing separation between $P_z(\phi)$ and $\overline{P}_z(\phi)$.
On the other hand, $P^{{\rm net}}_{2,y}$ shows an interesting non-monotonic dependence with the collision energy, which is insensitive to the choice of scenario.
Such non-trivial energy dependence is related to the net baryon density distribution correspondingly: $n_B(x,y)$ in transverse plane shows similar almond shape in all energies while its reaction plane profile $n_B(x,\eta_s)$ show different peak structure from baryon stopping. See our discussion in \cite{Fu:2022myl} for details.

%To understand these results better, we show the net baryon density distribution on transverse plane and reaction plane in Fig.~\ref{Fig:profile}, which are related to the $\mu_B$ gradients through EoS.
%In the transverse plane, the net baryon density shows similar almond shape that corresponding with the monotonic behavior of the longitudinal component $P^{{\rm net}}_{2,z}$.
%While for the reaction plane distribution, the profile shows a transition from double peak to single peak as a result of different baryon stopping effects, which explains the sign change of $P^{{\rm net}}_{2,y}$ with collision energy.

\section{Summary}

In this proceeding, we briefly summarize our recent study on the baryonic spin Hall effect (SHE) for local Lambda spin polarization. Based on MUSIC simulations with AMPT initial condition and spin Cooper-Fryer on the freeze-out surface, we found the baryonic SHE shows sizeable effect on $P_z(\phi)$ and $P_y(\phi)$ at RHIC BES energies, which leads to the local polarization separation between baryons and anti-baryons at 7.7 GeV. To isolate the baryonic SHE contribution, we predict the second Fourier coefficients of net spin polarization in longitudinal and transverse direction, which show non-trivial collision energy dependence and can be used as possible signals to detect baryonic SHE in future experiments.

\textit{Acknowledgments}~%
This work was supported in part by the NSFC under grant No.~12075007 and No.~11675004 (B.F. and H.S.), No.~12147173 (B.F.) and No.~11861131009, No.~12075098 (L.P.). 
Y.Y. thanks the support from the Strategic Priority Research Program of Chinese Academy of Sciences, Grant No. XDB34000000.
We also acknowledge the extensive computing resources provided by the Supercomputing Center of Chinese Academy of Science (SCCAS), Tianhe-1A from the National Supercomputing Center in Tianjin, China and the High-performance Computing Platform of Peking University.

\bibliographystyle{unsrtnt2}
\bibliography{ref}% Produces the bibliography via BibTeX.

\end{document}